\documentclass[a4paper,11pt]{article} 
\usepackage[latin1]{inputenc}
\usepackage{amsmath,amssymb,amsfonts}
\usepackage[authoryear]{natbib} 
\usepackage{graphicx}
\usepackage{setspace}

\newtheorem{theorem}{Theorem}[section]
\newtheorem{definition}{Definition}[section]

\newtheorem{proposition}[theorem]{Proposition}
 
\newtheorem{remark}[theorem]{Remark}
\usepackage{authblk}
\title{\bf A skew true INAR(1) process with application}
  \author[$\star$]{Wagner Barreto-Souza\thanks{Corresponding author. Email: wagnerbs85@gmail.com}}
    \author[$\ddagger$]{Marcelo Bourguignon\thanks{Email: m.p.bourguignon@gmail.com}}
\affil[$\star$]{Universidade de S\~ao Paulo, Departamento de Estat\'istica\\

  Butant\~a, 05508-090, S\~ao Paulo, SP, Brazil}
\affil[$\ddagger$]{Universidade Federal de Pernambuco, Departamento de Estat\'istica\\

  Cidade Universit\'aria, 50740-540, Recife, PE, Brazil} 
  
\begin{document}
\maketitle

\begin{abstract}
Integer-valued time series models have been a recurrent theme considered in many papers in the last three decades, but only a few of them have dealt with models on $\mathbb Z$ (that is, including both negative and positive integers). Our aim in this paper is to introduce a first-order integer-valued autoregressive process on $\mathbb Z$ with skew discrete Laplace marginals \citep{koin06}. For this, we define a new operator that acts on two independent latent processes, similarly as made by \citet{free10}. We derive some joint and conditional basic properties of the proposed process such as characteristic function, moments, higher-order moments and jumps. Estimators for the parameters of our model are proposed and their asymptotic normality are established. We run a Monte Carlo simulation to evaluate the finite-sample performance of these estimators. In order to illustrate the potentiality of our process, we apply it to a real data set about population increase rates.\\

\end{abstract}

\newpage

\section{Introduction}

Count time series models have been a recurrent theme considered in many papers in the last three decades. Pioneering works in this interesting theme are due to \cite{stevan1979}, \cite{mck1985}, \cite{alal87} and \cite{mck1988}. They introduced and studied count valued ARMA models with Poisson marginals. These models are constructed based on the binomial thinning operator. Issues such as inference and forecasting for Poisson ARMA models have been discussed by \cite{freemcc04a} and \cite{freemcc04b}. Asymptotic properties of estimators for a Poisson AR(1) model were established by \cite{freemcc05}.

\cite{ristic09} constructed and studied several properties of a stationary INAR(1) process with geometric marginals based on a negative binomial thinning operator; this model is named new geometric INAR(1) process (in short NGINAR). Further results on this model can be found in \cite{bak10}. The NGINAR(1) model is overdispersed and therefore it is an alternative to the Poisson AR(1) models. The literature about count time series models is too vast, so we recommend the readers to the papers above and the references contained therein. 

On the other hand, only a few papers have dealt with time series models on $\mathbb Z$ (that is, including both negative and positive integers). Such a models can arise naturally in practical situations. For example, it is frequent to encounter a non-stationary count time series. For instance, this happens with count series which are small in value and show a trend having relatively large fluctuation. To handle such a non-stationary series, the difference operator is commonly applied in the series to achieve stationarity. The differenced series may contain negative integer values and therefore the usual count models are not able to fit these data. With this in mind, some models were proposed in the literature. \cite{kp08} proposed an integer-valued autoregressive process on $\mathbb Z$ based on a signed binomial thinning. This process was recently generalized by \cite{zhang10}.

In a different approach of that considered by \cite{kp08}, \cite{free10} introduced a stationary AR(1) process on $\mathbb Z$ with symmetric Skellam marginals (which are distributed as a difference between two iid Poisson random variables); this model is named true INAR(1) process (in short TINAR). The idea of Freeland was to define a modified binomial thinning operator, which involves two iid latent Poisson AR(1) processes.

Our aim in this paper is to introduce a stationary INAR(1) process on $\mathbb Z$ with skew discrete Laplace marginals \citep{koin06}. We named this model by skew true INAR(1) process (in short STINAR). For this, we propose a modified version of the negative binomial thinning operator in a similar fashion as made by \cite{free10}. Here, our thinning operator acts on two independent but not necessarily identically distributed latent NGINAR(1) processes. The skew discrete Laplace (SDL) distributions (and other similar distributions) have a great importance in analysis of hydroclimatic episodes such as droughts, floods and El Niño; for instance, see the introduction section of \cite{koin06}. Due to these interesting applications of the SDL distribution, we think that our STINAR(1) process can also be of great interest in these areas when there is a temporal dependence.

We have some advantages of our model and the results obtained here with respect to that ones given in \cite{free10}. The main of them are:
\begin{itemize}
\item {\it Accommodation of skewness} (the TINAR(1) process is symmetric); 
\item {\it Mathematical simplicity of our model.} For instance, the probability and distribution functions of the skew discrete Laplace distribution have a simple form (see below) in contrast with the Skellam distribution which has associated probability function involving the modified Bessel function of the first kind.
\item {\it Full asymptotic behaviour of our proposed estimators.} As we will see later we establish the strong consistency and the asymptotic distributions (including the asymptotic covariance matrix) of the proposed estimators for the parameters of our model. In \cite{free10}, the asymptotic variance of the estimator proposed for the parameter related to the counting series is not obtained explicitly. 
\end{itemize}

The present paper is organized in the following way. In Section \ref{stinar(1)process} we introduce our skew true INAR(1) process. We also obtain some statistical properties such as joint moments and joint and conditional characteristic functions among others. In Section \ref{higherordermomentsandjumps}, we obtain joint higher-order moments and present some properties of the jumps for the STINAR(1) process. Estimators for the parameters of our model and their asymptotic normality are presented in Section \ref{estimationandinference}. In Section \ref{simulationissues} we present some simulation results in order to evaluate the finite-sample performance of the proposed estimators. In Section \ref{application} we illustrate the potentiality of our model by applying it to a real data set about population increase rates. 

\section{STINAR(1) process}\label{stinar(1)process}

In this section we introduce a stationary first-order integer-valued autoregressive process on $\mathbb Z$ with skew discrete Laplace marginals, named skew true INAR(1) model (in short STINAR). For this, we first introduce some notation and the NGINAR(1) model by \cite{ristic09}.

Let $\alpha\in[0,1)$ and $``\ast"$ be the negative binomial thinning operator \citep{ristic09}, which is defined by
\begin{eqnarray*}\label{nbthin}
\alpha \ast X=\sum_{i=1}^XW_{iX},
\end{eqnarray*}
for $X\in\mathbb{N}$ and $\alpha \ast 0=0$, where $\{W_{ij}\}_{i,j\in\mathbb N}$ is a sequence of iid random variables following a geometric distribution on $\mathbb N$ with mean $\alpha$. 

\begin{definition}\label{nginarp} (NGINAR(1) process)
Let $\{X_t\}_{t\in \mathbb{N}}$ be a stationary process having geometric marginals with probability function assuming the form $P(X_t=x)= \mu^{x}/(1+\mu)^{x+1}$, where $x\in\mathbb N$ and $\mu>0$. The NGINAR(1) process is defined by
\begin{eqnarray*}\label{nginar}
X_t=\alpha\ast X_{t-1}+\epsilon_t, \quad t \in \mathbb{N}^*,
\end{eqnarray*}
where $\{\epsilon_t\}_{t\in\mathbb N^*}$ is a sequence of iid random variables independent of $\{W_{ij}\}$, and $\epsilon_t$ and $X_{t-l}$ are independent for all $l\geq1$, with $\mathbb N^*=\mathbb N\backslash\{0\}$.
\end{definition}

\citet{ristic09} showed that the probability function of $\epsilon_t$ is given by
\begin{equation}\label{epsilon}
P(\epsilon_t=l) = \left(1 - \frac{\alpha\mu}{\mu-\alpha}\right)\frac{\mu^l}{(1+\mu)^{l+1}}
+ \frac{\alpha\mu}{\mu-\alpha}\frac{\alpha^l}{(1+\alpha)^{l+1}}, \quad l\in\mathbb N^*,
\end{equation}
that is, the random variable $\epsilon_t$ is a mixture of two independent random variables that follow geometric distributions with means $\mu$ and $\alpha$.

We now present briefly the skew discrete Laplace (SDL) distribution studied in \cite{koin06}, which will be the marginal of our process. A discrete random variable $Z$ following a SDL distribution with parameters $\mu_1>0$ and $\mu_2>0$ has probability and distribution functions given by
\begin{eqnarray}\label{prob}
p(k;\mu_1,\mu_2)\equiv P(Z=k)=\frac{1}{1+\mu_1+\mu_2}\left\{\begin{array}{ll} 
\bigg(\dfrac{\mu_1}{1+\mu_1}\bigg)^k,&k = 0, 1, 2, \ldots \\
\,\bigg(\dfrac{\mu_2}{1+\mu_2}\bigg)^{|k|},&k = 0, -1, -2, \ldots\\
\end{array}\right.
\end{eqnarray}
and
\begin{eqnarray*}
P(Z\leq k)=\left\{\begin{array}{ll}1-\dfrac{(1+\mu_1+\mu_2)^{-1}\mu_1^k}{(1+\mu_1)^{k+1}},&k = 0, 1, 2, \ldots \\
\dfrac{(1+\mu_2)^{k+1}}{(1+\mu_1+\mu_2)\mu_2^k},&k = 0, -1, -2, \ldots\\
\end{array}\right.,
\end{eqnarray*}
respectively. The SDL distribution shares many of the properties of the skew (continuous) Laplace distribution such as infinitely divisibility, closure under geometric summation and a maximum entropy property. Moreover, a random variable following this distribution can be stochastically represented as a difference between two independent but not necessarily identically distributed geometric random variables. For more detail and other properties, see \cite{koin06}.

With the notations and definitions above, we are ready to introduce our STINAR(1) process.

\begin{definition}(STINAR(1) process) Let $\{X_t\}_{t\in\mathbb N}$ and $\{Y_t\}_{t\in\mathbb N}$ be two independent NGINAR(1) processes with geometric marginals with mean $\mu_1>0$ and $\mu_2>0$ (respectively) and common parameter $\alpha>0$ related to the counting series, as presented in Definition \ref{nginarp}. More specifically, we define
$$X_t=\alpha\ast X_{t-1}+\epsilon_t, \quad t \in \mathbb{N}^*$$
and
$$Y_t=\alpha\ast Y_{t-1}+\nu_t, \quad t \in \mathbb{N}^*.$$
The sequences $\{\epsilon_t\}_{t \in \mathbb{N}^*}$ and $\{\nu_t\}_{t \in \mathbb{N}^*}$ are the innovations of the processes $\{X_t\}_{t\in\mathbb N}$ and $\{Y_t\}_{t\in\mathbb N}$ (respectively) and are defined as that one of Definition \ref{nginarp}. Let $\{Z_t\}_{t\in\mathbb N^*}$ be a sequence of random variables following a common skew discrete Laplace distribution with parameters $\mu_1$ and $\mu_2$, and define $\varepsilon_t=\epsilon_t-\nu_t$, for $t\in\mathbb N^*$. Then, we define our modified negative binomial thinning operator $``\odot"$ by
\begin{eqnarray*}\label{op}
\alpha\odot Z_{t-1}\stackrel{d}{=}\alpha\ast X_{t-1}-\alpha\ast Y_{t-1},
\end{eqnarray*}
for $t\in\mathbb N^*$. With this, we define completely our STINAR(1) process $\{Z_t\}_{t\in\mathbb N}$ by
\begin{eqnarray*}
Z_t=\alpha\odot Z_{t-1}+\varepsilon_t,
\end{eqnarray*}
for $t\in\mathbb N^*$.
\end{definition}

\begin{remark}\label{restriction}
From the results of \cite{ristic09}, we have that the STINAR(1) process is well-defined for $\alpha\leq \min\{\mu_1/(1+\mu_1),\mu_2/(1+\mu_2)\}$. With this, it is possible to find the distribution of $\varepsilon_t$, as will be discussed below. From now on, we consider the STINAR(1) process with this restriction on $\alpha$.
\end{remark}

\begin{remark}
If $\mu_1=\mu_2$, $\{Z_t\}_{t\in\mathbb N}$ is a symmetric true INAR(1) process with symmetric discrete Laplace marginals \citep{inko06}.
\end{remark}

\begin{remark}
If $\mu_2=0$, $\{Z_t\}_{t\in\mathbb N}$ is the NGINAR(1) process proposed by \cite{ristic09}.
\end{remark}

\begin{figure}[!h]
\centering \caption{Plots of simulated trajectories of the STINAR(1) process for $\alpha=0.1, 0.4, 0.7, 0.8$ and $(\mu_1,\mu_2)=(5,5)$.}
		 \includegraphics[width=0.45\textwidth]{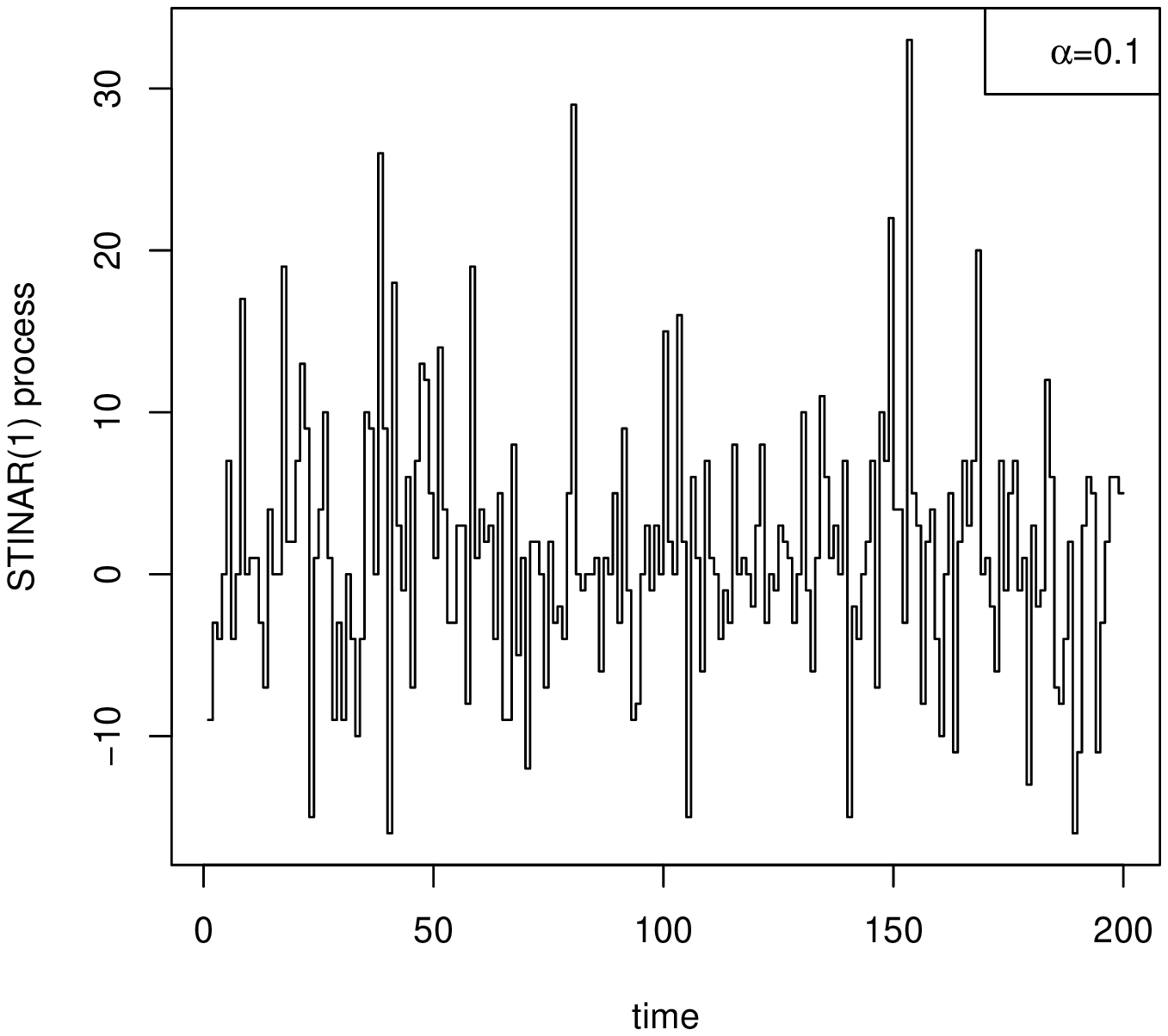}\includegraphics[width=0.45\textwidth]{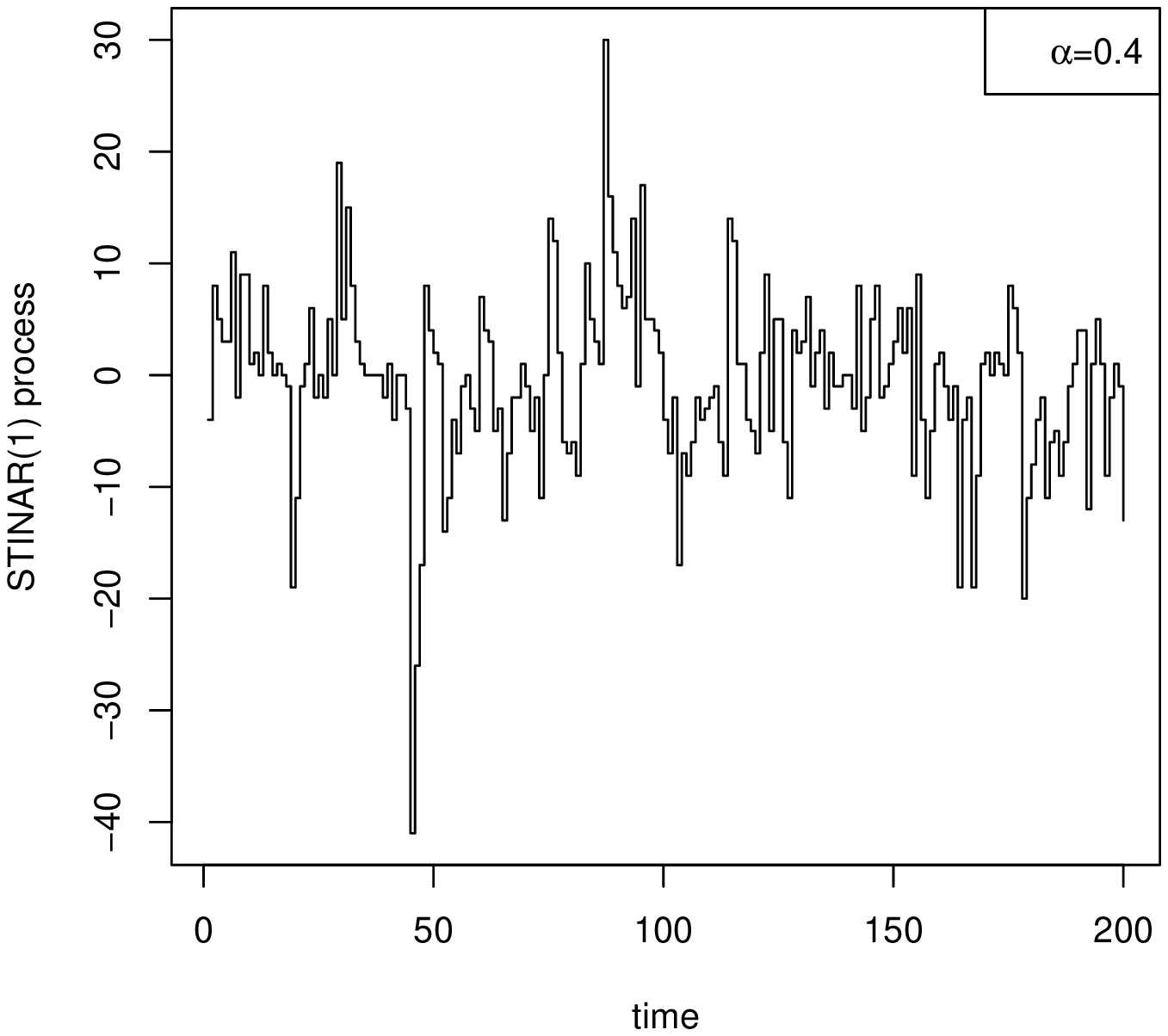}
		 \includegraphics[width=0.45\textwidth]{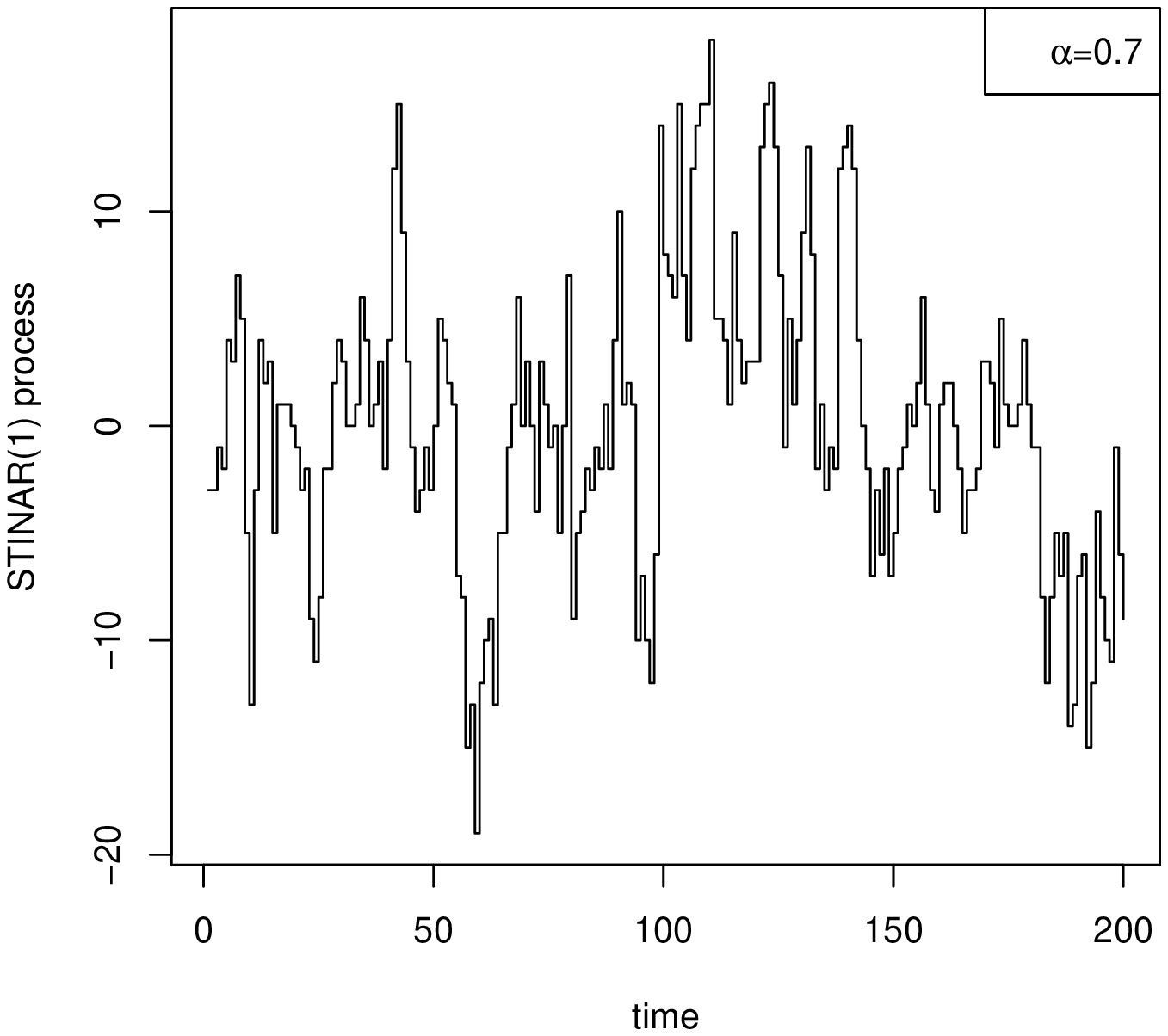}\includegraphics[width=0.45\textwidth]{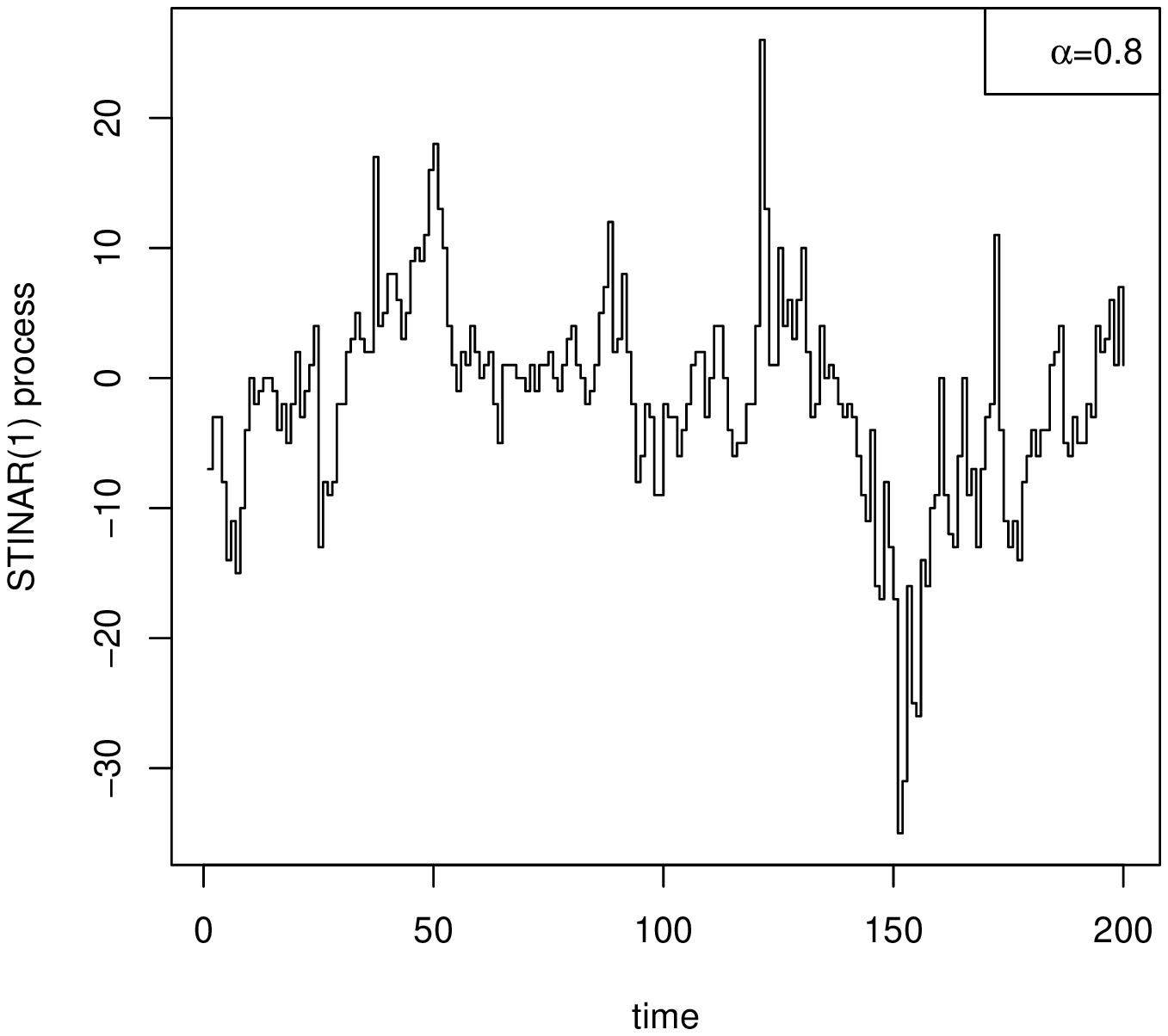}
\label{fig1}
\end{figure}

\begin{figure}[!h]
\centering \caption{Plots of simulated trajectories of the STINAR(1) process for $\alpha=0.1, 0.4, 0.7, 0.8$ and $(\mu_1,\mu_2)=(10,5)$.}
		 \includegraphics[width=0.45\textwidth]{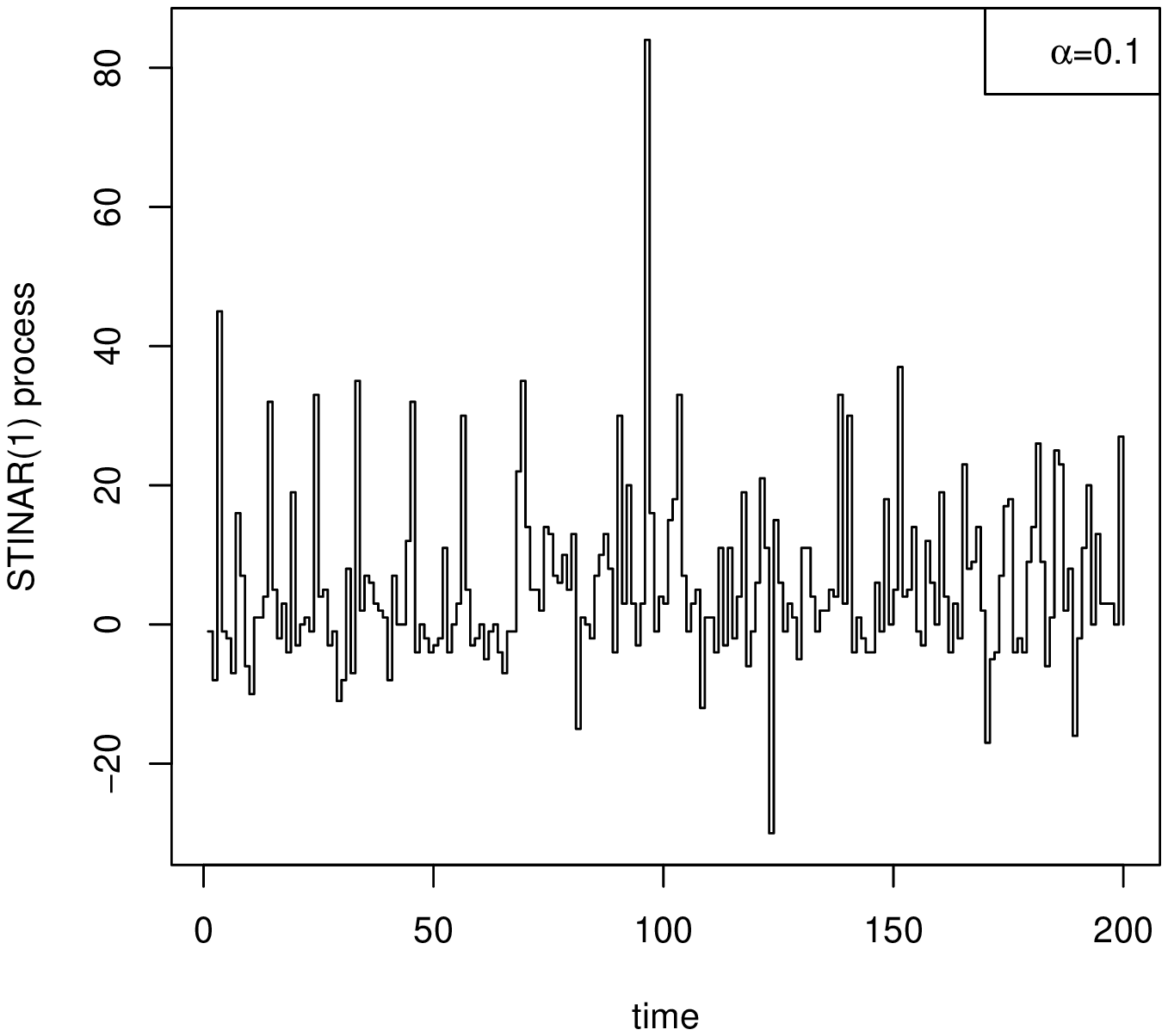}\includegraphics[width=0.45\textwidth]{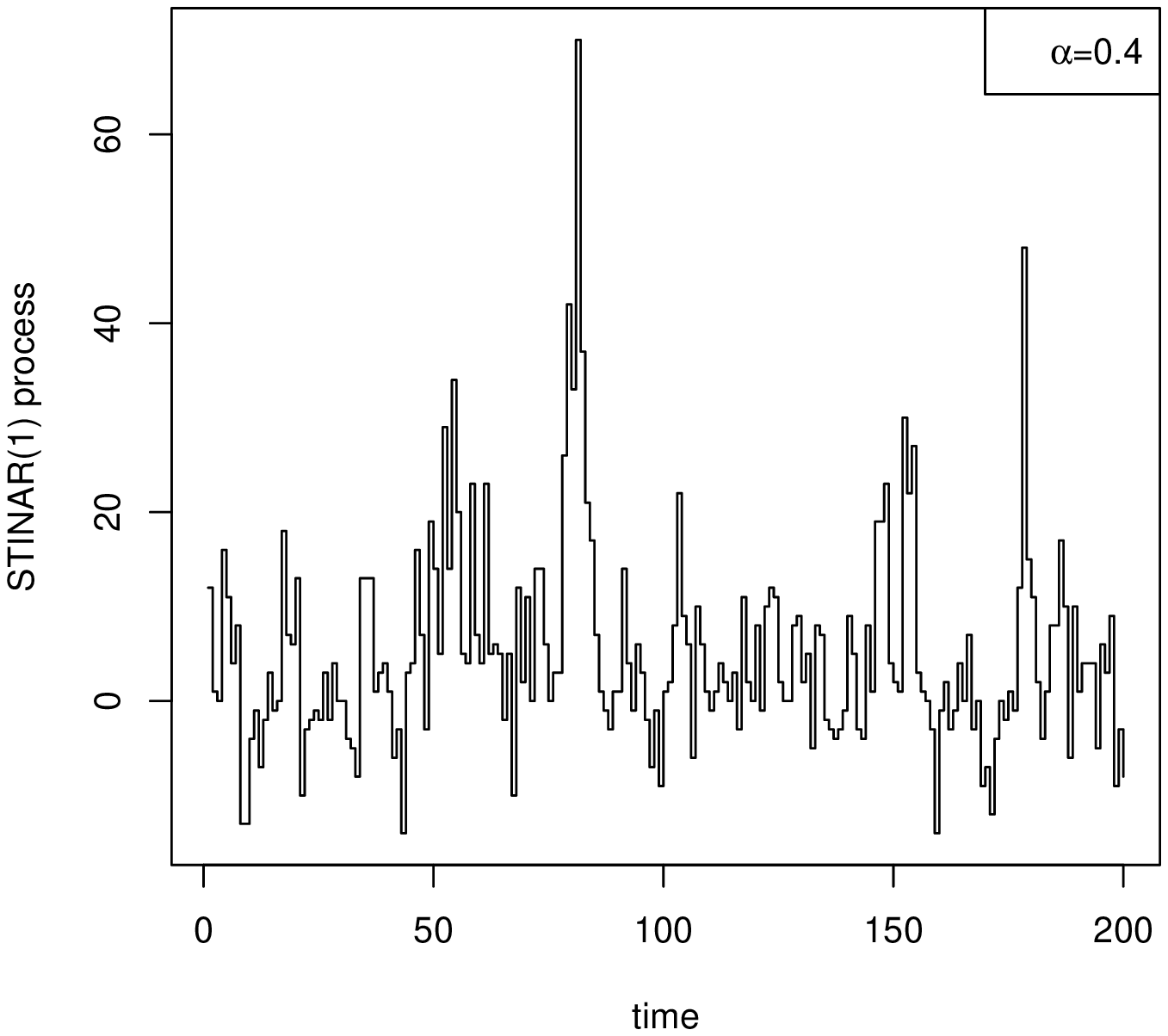}
		 \includegraphics[width=0.45\textwidth]{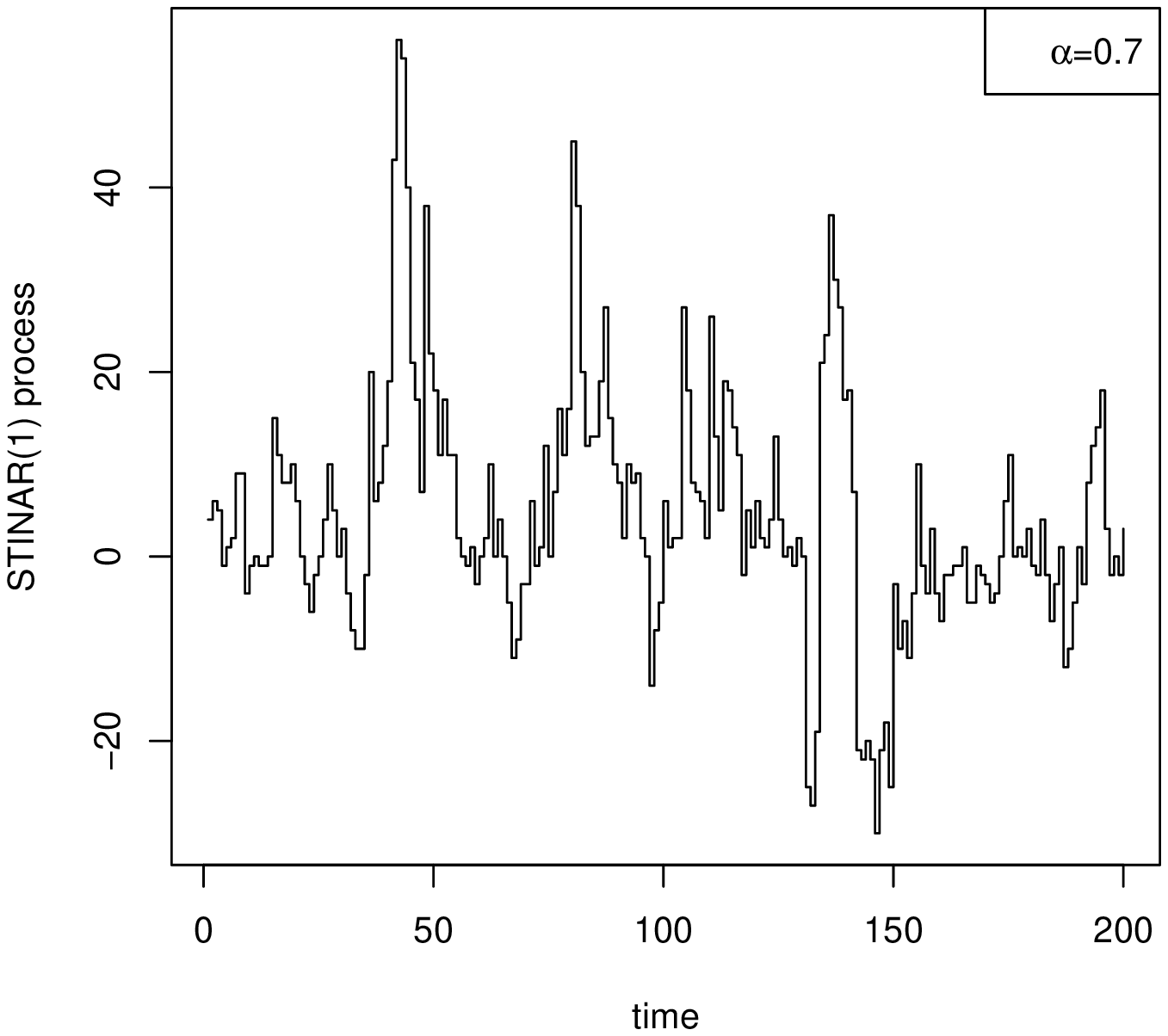}\includegraphics[width=0.45\textwidth]{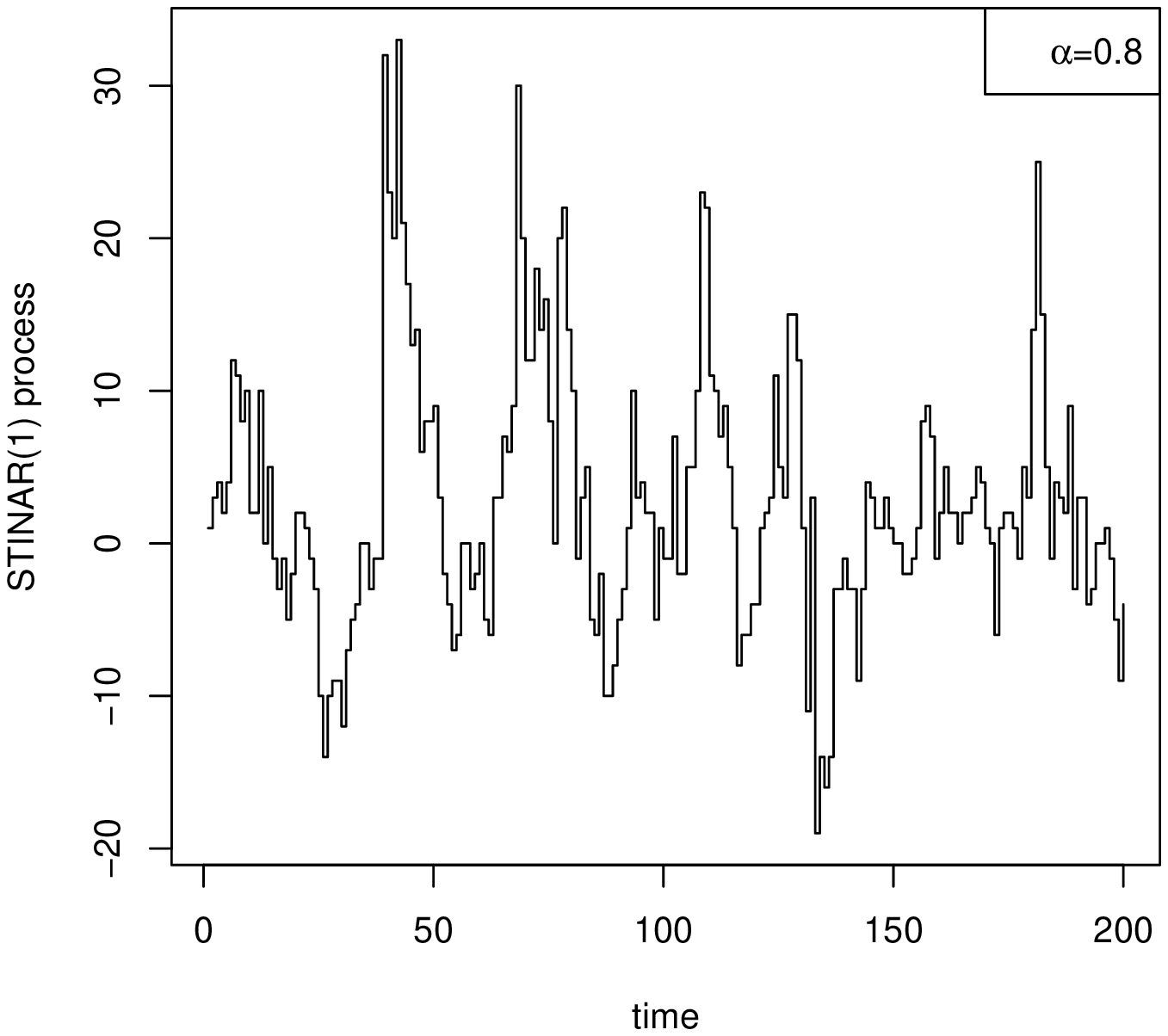}
\label{fig2}
\end{figure}

In Figures \ref{fig1} and \ref{fig2} we present some simulated trajectories of the STINAR(1) process for $\alpha=0.1, 0.4, 0.7, 0.8$ and $(\mu_1,\mu_2)=(5,5)$ and $(\mu_1,\mu_2)=(10,5)$, respectively.\\

Marginal properties of our process can be obtained directly from the results given in \cite{koin06}. For example, the characteristic function of $Z_t$, denoted by $\phi(s)\equiv E(\exp(isZ_t))$ (with $i=\sqrt{-1}$), is given by
\begin{eqnarray*}
\phi(s)=\{[1+\mu_1(1-e^{is})][1+\mu_2(1-e^{-is})]\}^{-1}, \quad s\in\mathbb R.
\end{eqnarray*}

The moments and absolute moments of $Z_t$ are given by
\begin{eqnarray*}\label{mom}
E(Z_t^k)=\frac{(1+\mu_1)(1+\mu_2)}{1+\mu_1+\mu_2}\sum_{j=1}^kj!S(k,j)\bigg(\frac{\mu_1^j}{1+\mu_2}+(-1)^k\frac{\mu_2^j}{1+\mu_1}\bigg)
\end{eqnarray*}
and
\begin{eqnarray}\label{absolutemoments}
E(|Z_t|^k)=\frac{(1+\mu_1)(1+\mu_2)}{1+\mu_1+\mu_2}\sum_{j=1}^kj!S(k,j)\bigg(\frac{\mu_1^j}{1+\mu_2}+\frac{\mu_2^j}{1+\mu_1}\bigg),
\end{eqnarray}
respectively, where $S(k,j)=j!^{-1}\sum_{i=0}^{j-1}(-1)^i\binom{j}{i}(j-i)^k$ is the Stirling number of second kind.
In particular, the expected value, variance and first absolute moment of $Z_t$ are given by
\begin{eqnarray}\label{meanvarZ}
\mu\equiv E(Z_t) = \mu_1 - \mu_2, \quad \sigma^2 \equiv \textrm{Var}(Z_t) =  \mu_1(1 + \mu_1) + \mu_2(1 + \mu_2)
\end{eqnarray}
and
\begin{equation*}
E(|Z_t|) = \frac{\mu_1(1 + \mu_1) + \mu_2(1 + \mu_2)}{1 + \mu_1 + \mu_2},
\end{equation*}respectively.

With the restriction given in the Remark \ref{restriction} and using the definition of $\varepsilon_t$ and the result (\ref{epsilon}), we obtain that the probability function of $\varepsilon_t$ can be expressed by
\begin{eqnarray*}
P(\varepsilon_t=k)=\beta_1\,p(k;\mu_1,\mu_2)+\beta_2\,p(k;\mu_1,\alpha)+\beta_3\,p(k;\alpha,\mu_2)+(1-\beta_1-\beta_2-\beta_3)\,p(k;\alpha,\alpha),
\end{eqnarray*}
for $k\in\mathbb Z$, where $p(k;\cdot,\cdot)$ is defined in (\ref{prob}) and
\begin{eqnarray*}
\beta_1=\bigg(1-\frac{\alpha\mu_1}{\mu_1-\alpha}\bigg)\bigg(1-\frac{\alpha\mu_2}{\mu_2-\alpha}\bigg),\,\,\,
\beta_2=\bigg(1-\frac{\alpha\mu_1}{\mu_1-\alpha}\bigg)\frac{\alpha\mu_2}{\mu_2-\alpha},\,\,\,
\beta_3=\frac{\alpha\mu_1}{\mu_1-\alpha}\bigg(1-\frac{\alpha\mu_2}{\mu_2-\alpha}\bigg).
\end{eqnarray*}

That is, the random variable $\varepsilon_t$ is distributed as a mixture of skew discrete Laplace random variables. Using this, it is straighforward to obtain that the characteristic function of $\varepsilon_t$, denoted by $\phi_\varepsilon(s)=E(\exp(is\varepsilon_t))$, is given by
\begin{eqnarray}\label{phiep}
\phi_\varepsilon(s)=\frac{[1+\alpha(1+\mu_1)(1-e^{is})][1+\alpha(1+\mu_2)(1-e^{-is})]}{[1+\mu_1(1-e^{is})][1+\alpha(1-e^{is})][1+\mu_2(1-e^{-is})][1+\alpha(1-e^{-is})]},
\end{eqnarray}
for $s\in\mathbb R$. We have that the two first cumulants of $\varepsilon_t$ are given by
\begin{eqnarray}\label{meanepsilon}
\mu_\epsilon&\equiv& E(\varepsilon_t)=(1-\alpha)(\mu_1-\mu_2),\\
\label{varepsilon}
\sigma^2_\epsilon&\equiv&\textrm{Var}(\varepsilon_t)=(1+\alpha)\{\mu_1[(1-\alpha)(1+\mu_1)-\alpha]+\mu_2[(1-\alpha)(1+\mu_2)-\alpha]\}.
\end{eqnarray}

From our definition, we have that the STINAR(1) process can be seen as a difference between two independent NGINAR(1) processes, that is, $Z_t\stackrel{d}{=}X_t-Y_t$, with $X_t$ and $Y_t$ as in Definition \ref{nginarp}. From this, we can obtain some properties for our process from the properties of the NGINAR(1) process. The following proposition states some results that follows from this fact.

\begin{proposition} The following results are valid for the STINAR(1) model:\\
(i) It is markovian, stationary and ergodic;\\
(ii) The autocorrelation $\rho(k)=\textrm{Corr}(Z_{t},Z_{t-k})$ is given by
$\rho(k)=\alpha^{k}$, for $k\in\mathbb N^*$;\\
(iii) The spectral density function $f(\omega)\equiv(2\pi)^{-1}\sum_{k\in\mathbb Z}\rho(k)e^{-ik\omega}$ reduces to
\begin{eqnarray*}
f(\omega)=\frac{(1-\alpha^2)}{2\pi}\frac{\mu_1(1+\mu_1) + \mu_2(1+\mu_2)}{1+\alpha^2-2\alpha\cos(\omega)},\quad \omega \in (-\pi,\pi].
\end{eqnarray*}
\end{proposition}

\begin{remark}
From the proposition above, we see that our STINAR(1) process has positive autocorrelation. It is possible to define an INAR(1) process having negative autocorrelation and with SDL marginals. For this, we follow the idea of \cite{free10} and define $$Z_t\stackrel{d}{=}\left\{\begin{array}{ll} 
X_t-Y_t,&t=0,2,4, \ldots \\
Y_t-X_t,&t=1,3,5, \ldots \\
\end{array}\right.,$$
with $\varepsilon_t=\epsilon_t-\eta_t$ for $t=0,2,4,\ldots$ and $\varepsilon_t=\eta_t-\epsilon_t$ for $t=1,3,5,\ldots$. In this case, it can be shown that $\rho(k)=\textrm{Corr}(Z_{t},Z_{t-k})=(-\alpha)^k$, for $k\in\mathbb N^*$. The results for this process follow in a similar fashion to that ones related to the STINAR(1) model considered along this paper.
\end{remark}

We now obtain the conditional characteristic function of $Z_t|Z_{t-1}$ and the joint characteristic function of $(Z_t,Z_{t-1})$. The expressions of the associated probability functions are cumbersome and therefore omitted here.

Denote by $N_r$ a random variable following a negative binomial distribution with parameters $r\geq0$ and $\beta>0$ with probability function assuming the form
$$P(N_k=k)=\binom{k+r-1}{k}\bigg(\frac{1}{1+\beta}\bigg)^r\bigg(\frac{\beta}{1+\beta}\bigg)^k,$$
for $k\in\mathbb N^*$. The associated characteristic function $\phi_{N_r}(s)\equiv E(\exp(isN_r))$ is given by
$$\phi_{N_r}(s)=[1+\beta(1-e^{is})]^{-r},$$
for $s\in\mathbb R$. Let $z$ be a non-negative integer. Using the stochastic representation of $Z_t$ and the characteristic function above, we obtain that
\begin{eqnarray*}
E(\exp(isZ_t)|Z_{t-1}=z)&=&\frac{\phi_{\varepsilon}(s)}{P(Z_{t-1}=z)}\sum_{y=0}^\infty P(X_{t-1}=z+y,Y_{t-1}=y)\phi_{N_{z+y}}(s)\phi_{N_y}(-s)\nonumber\\
&=&\frac{(1+\mu_1+\mu_2)\phi_{\varepsilon}(s)[1+\alpha(1-e^{is})]^{1-z}[1+\alpha(1-e^{-is})]}{(1+\mu_1)(1+\mu_2)[1+\alpha(1-e^{is})][1+\alpha(1-e^{-is})]-\mu_1\mu_2}, 
\end{eqnarray*}
for $s\in\mathbb R$, where $\phi_\varepsilon(\cdot)$ is the characteristic function of $\varepsilon_t$ given in (\ref{phiep}).
For $z<0$ integer, it can be shown in a similar way that the conditional characteristic function $E(\exp(isZ_t)|Z_{t-1}=z)$ can be expressed by
\begin{eqnarray*}
E(\exp(isZ_t)|Z_{t-1}=z)=\frac{(1+\mu_1+\mu_2)\phi_{\varepsilon}(s)[1+\alpha(1-e^{is})][1+\alpha(1-e^{-is})]^{1+z}}{(1+\mu_1)(1+\mu_2)[1+\alpha(1-e^{is})][1+\alpha(1-e^{-is})]-\mu_1\mu_2}, \,\, s\in\mathbb R.
\end{eqnarray*}

Hence, we obtain in particular that the mean and variance of $Z_t$ given $Z_{t-1}$ are given by
\begin{eqnarray*}
E(Z_t|Z_{t-1}=z)=\mu_\varepsilon+\alpha z
\end{eqnarray*}
and
\begin{eqnarray*}
\textrm{Var}(Z_t|Z_{t-1}=z)=\sigma^2_\varepsilon+\alpha(1+\alpha)|z|+\frac{2\alpha(1+\alpha)\mu_1\mu_2}{1+\mu_1+\mu_2},
\end{eqnarray*}
respectively, for $z\in\mathbb Z$, with $\mu_\epsilon$ and $\sigma^2_\epsilon$ given in (\ref{meanepsilon}) and (\ref{varepsilon}), respectively.

We now obtain an expression for the joint characteristic function of $(Z_t,Z_{t-1})$, which we denote by $\phi(s,u)=E(\exp(isZ_t+iuZ_{t-1}))$, for $u,s\in\mathbb R$. We have that $$\phi(s,u)=E\{\exp(iuZ_{t-1})E(\exp(isZ_t)|Z_{t-1})\}.$$

Hence, it can be shown that the double expectation in the right side of the above equation can be expressed by
$$E\{\exp(iuZ_{t-1})E(\exp(isZ_t)|Z_{t-1})\}=(1+\mu_1+\mu_2)\varphi(t)\{E({s^*}^{Z_t}1\{Z_t\geq0\})+E({s^\dag}^{|Z_t|}1\{Z_t<0\})\},$$
where
\begin{eqnarray}\label{varphiaux}
\varphi(s)=\frac{\phi_\varepsilon(s)[1+\alpha(1-e^{is})][1+\alpha(1-e^{-is})]}{(1+\mu_1)(1+\mu_2)[1+\alpha(1-e^{is})][1+\alpha(1-e^{-is})]-\mu_1\mu_2},
\end{eqnarray}
$s^*=e^{iu}[1+\alpha(1-e^{is})]^{-1}$ and $s^\dag=e^{iu}[1+\alpha(1-e^{-is})]^{-1}$. It can be checked that
$$E({s^*}^{Z_t}1\{Z_t\geq0\})=\frac{1+\mu_1}{1+\mu_1+\mu_2}\frac{1+\alpha(1-e^{is})}{(1+\mu_1)[1+\alpha(1-e^{is})]-e^{iu}\mu_1}$$
and
$$E({s^\dag}^{|Z_t|}1\{Z_t<0\})=\frac{\mu_2}{1+\mu_1+\mu_2}\frac{e^{iu}}{(1+\mu_2)[1+\alpha(1-e^{-is})]-e^{iu}\mu_2}.$$

With the results above we obtain that the joint characteristic function of $(Z_t,Z_{t-1})$ can be expressed by
\begin{eqnarray}\label{jointcharact}
\phi(s,u)=\varphi(s)\left\{\frac{(1+\mu_1)[1+\alpha(1-e^{is})]}{(1+\mu_1)[1+\alpha(1-e^{is})]-e^{iu}\mu_1}+\frac{\mu_2e^{iu}}{(1+\mu_2)[1+\alpha(1-e^{-is})]-e^{iu}\mu_2}\right\},
\end{eqnarray}
where $\varphi(\cdot)$ is defined in (\ref{varphiaux}).

\section{Higher-order moments and jumps}\label{higherordermomentsandjumps}

This section is devoted to find some additional statistical measures than those given in the previous section. We here obtain joint higher-order moments for our process $Z_t$ and study the jump process, which is defined by $J_t\equiv Z_t-Z_{t-1}$, for $t\in\mathbb N^*$. We use the following notation for the higher-order moments:
$$\mu(s_1,\ldots,s_{r-1})\equiv E(Z_tZ_{t+s_1}\ldots Z_{t+s_{r-1}}),$$
with $0\leq s_1\leq \ldots\leq s_{r-1}$ and $r\in\mathbb N$. In the following proposition, we present the second and third-order joint moments of $Z_t$ (the first-order moment was presented in the previous section). This result can be obtained by using the stochastic representation of $Z_t$ (that is, $Z_t$ as a difference between two independent NGINAR(1) processes) and the results given in Theorem 1 from \citet{bak10}.

\begin{proposition}
The second-order and third-order joint moments of the STINAR(1) process are given by
\begin{eqnarray*}
\mu(s)=\alpha^s[\mu_1(1+\mu_1)+\mu_2(1+\mu_2)]+(\mu_1-\mu_2)^2,
\end{eqnarray*}
\begin{eqnarray*}
\mu(0,s)&=&2(2+\alpha^s)\mu_1\mu_2(\mu_2-\mu_1)+\mu_1^2(1+2\mu_1)-\mu_2^2(1+2\mu_2)+\\
&&\alpha^s[\mu_1(1+5\mu_1+4\mu_1^2)-\mu_2(1+5\mu_2+4\mu_2^2)],
\end{eqnarray*}
\begin{eqnarray*}
\mu(s,s)&=&2(2+\alpha^s)\mu_1\mu_2(\mu_2-\mu_1)+\mu_1^2(1+2\mu_1)-\mu_2^2(1+2\mu_2)+\\
&& \frac{2\alpha^{s+1}}{1-\alpha}[\mu_2(1+\mu_2)^2-\mu_1(1+\mu_1)^2]+\frac{2\alpha^{2s}}{1-\alpha}[\mu_1^2(1+\mu_1)-\mu_2^2(1+\mu_2)]+\\
&&\frac{\alpha^{s+1}}{1-\alpha}[\mu_1(1+\mu_1)(1-2\mu_1)-\mu_2(1+\mu_2)(1-2\mu_2)]+\\
&&\frac{\alpha^s}{1-\alpha}[\mu_1(1+\mu_1)(1+2\mu_1)-\mu_2(1+\mu_2)(1+2\mu_2)],
\end{eqnarray*}
\begin{eqnarray*}
\mu(s,u)&=&-\mu_2[3\mu_1^2+\mu_1(1+\mu_1)(\alpha^s+\alpha^u+\alpha^{u-s})]+
\mu_1[3\mu_2^2+\mu_2(1+\mu_2)(\alpha^s+\alpha^u+\alpha^{u-s})]\\
&&\frac{\alpha^{u-s}}{1-\alpha}\{2\alpha^{2s+1}[\mu_2(1+\mu_2)^2-\mu_1(1+\mu_1)^2]+2\alpha^{2s}[\mu_1^2(1+\mu_1)^2-\mu_2^2(1+\mu_2)]+\\
&&\alpha^{s+1}[\mu_1^2(1+\mu_1)(1-2\mu_1)-\mu_2(1+\mu_2)(1-2\mu_2)]+\\
&&\alpha^s [\mu_1^2(1+\mu_1)(1+2\mu_1)-\mu_2(1+\mu_2)(1+2\mu_2)]\}+\\
&&\alpha^{u-s}[\mu_1^2(1+2\mu_1)-\mu_2^2(1+2\mu_2)]+\\
&&(1-\alpha^{u-s})\{\alpha^s[\mu_1^2(1+\mu_1)-\mu_2^2(1+\mu_2)]+\mu_1^3-\mu_2^3\},\quad s<u.
\end{eqnarray*}
\end{proposition}

We now focus on the properties of the jump process $J_t=Z_t-Z_{t-1}$, $t\in\mathbb N^*$. Jump processes have been considered and studied in the literature due to applications in checking the adequacy of the fitted model. Further, they have been used to construct control charts to detect changes in the serial dependence structure, as proposed by \cite{weib09b}. For instance, jumps in the Poisson and binomial count processes were investigated by \citet{weib08,weib09a} and \cite{weib09b}, respectively.

Taking $u=-s$ in (\ref{jointcharact}), we obtain that the characteristic function of $J_t$ can be expressed by
\begin{eqnarray*}
E(\exp(isJ_t))=\varphi(s)\left\{\frac{(1+\mu_1)[1+\alpha(1-e^{is})]}{(1+\mu_1)[1+\alpha(1-e^{is})]-e^{-is}\mu_1}+\frac{\mu_2e^{-is}}{(1+\mu_2)[1+\alpha(1-e^{-is})]-e^{-is}\mu_2}\right\},
\end{eqnarray*}
where $\varphi(\cdot)$ is given in (\ref{varphiaux}). We now present the first three moments of $J_t$. These results can be obtained directly or using the characteristic function above.
\begin{proposition}\label{propmomjumps}
The first three moments of the jump process $J_t$ are $E(J_t)=0$, $E(J_t^2)=2(1-\alpha)[\mu_1(1+\mu_1)+\mu_2(1+\mu_2)]$ and
$E(J_t^3)=3\alpha[\mu_1(1+5\mu_1+4\mu_1^2)-\mu_2(1+5\mu_2+4\mu_2^2)]-6\alpha^3(1-\alpha)^{-1}[\mu_2(1+\mu_2)^2-\mu_1(1+\mu_1)^2]
-3\alpha^2(1-\alpha)^{-1}[\mu_1(1+\mu_1)-\mu_2(1+\mu_2)]-3\alpha(1-\alpha)^{-1}[\mu_1(1+\mu_1)(1+2\mu_1)-\mu_2(1+\mu_2)(1+2\mu_2)]$.
\end{proposition}

We also obtain the autocorrelation function of $J_t$, which is denoted here by $\rho_J(\cdot)$. Using the autocorrelation function $\rho(\cdot)$ of $Z_t$, it can be shown that
$$\rho_J(k)=2\rho(k)-\rho(k+1)-\rho(k-1)=-\alpha^{k-1}(1-\alpha)^2,$$
for $k\in\mathbb N^*$. Note that the autocorrelation function of $J_t$ is always negative.

\section{Estimation and inference}\label{estimationandinference}

We here propose estimators for the parameters of our process and find their asymptotic distributions. We do not here consider estimation by maximum likelihood since the likelihood for our model is cumbersome to work with. Let $n$ be the sample size of the time series $Z_t$. We start proposing a estimator for the parameter $\alpha$ based on the conditional least square method. In this case, the function to be minimized is given by
$$Q_{n}(\alpha,\mu) = \sum_{t=2}^{n}(Z_t - \alpha Z_{t-1} - (1-\alpha)\mu)^2.$$

Note that this method does not provide estimators for $\mu_1$ and $\mu_2$, but only for $\alpha$ and $\mu=\mu_1-\mu_2$.
For the symmetric case $\mu_1=\mu_2$ ($\mu=0$), we obtain that the estimator $\widehat\alpha$ of $\alpha$ becomes
\begin{eqnarray*}\label{cls1}
\widehat{\alpha} =
\frac{\sum_{t=2}^{n}Z_{t}Z_{t-1}}{\sum_{t=2}^{n}Z_{t-1}^{2}}.
\end{eqnarray*}

Under the non-symmetric case $\mu_1\neq\mu_2$ ($\mu\neq0$), we obtain that the estimator $\widehat\alpha$ of $\alpha$ is given by
\begin{eqnarray*}\label{cls1}
\widehat{\alpha} =
\frac{(n-1)\sum_{t=2}^{n}Z_{t}Z_{t-1}-\sum_{t=2}^{n}Z_t\sum_{t=2}^{n}Z_{t-1}}{(n-1)\sum_{t=2}^{n}Z_{t-1}^{2}
- \left(\sum_{t=2}^{n}Z_{t-1}\right)^2}.
\end{eqnarray*}

In the next proposition we establish the strong consistency and the asymptotic distribution of $\widehat\alpha$, which is valid in both symmetric and non-symmetric cases.

\begin{proposition}\label{propalphaad}
The estimator $\widehat\alpha$ is strongly consistent for $\alpha$ and satisfy the asymptotic normality
\begin{eqnarray}\label{an}
\sqrt{n}(\widehat\alpha-\alpha)\stackrel{d}{\longrightarrow}{\mbox N}(0,\nu^2),
\end{eqnarray}
as $n\rightarrow\infty$, with $$\nu^2=\frac{1}{\sigma^2}\bigg(\sigma^2_\varepsilon+\frac{2\alpha(1+\alpha)\mu_1\mu_2}{1+\mu_1+\mu_2}\bigg)+\frac{\alpha(\alpha+1)}{\sigma^4}[E(|Z|^3)-2\mu E(sgn(Z)Z^2)+\mu^2E(|Z|)],$$
where $sgn(Z)=1$ if $Z\geq0$ and $sgn(Z)=-1$ if $Z<0$, $\sigma^4\equiv(\sigma^2)^2$ and $Z$ is a random variable following a SDL distribution with parameters $\mu_1$ and $\mu_2$. The variances $\sigma^2$ and $\sigma^2_\epsilon$ are given in (\ref{meanvarZ}) and (\ref{varepsilon}), respectively.  
\end{proposition}

\begin{remark}
The first and third moments of $|Z|$ involved in the asymptotic variance of $\widehat\alpha$ can be obtained from (\ref{absolutemoments}). The expected value of $\mbox{sgn}(Z)Z^2$ is given by $$E(\mbox{sgn}(Z)Z^2)=\frac{\mu_1(1 + \mu_1)(1 + 2\mu_1) - \mu_2(1 + \mu_2)(1 + 2\mu_2)}{1 + \mu_1 + \mu_2}.$$
\end{remark}

\noindent {\bf Proof of Proposition \ref{propalphaad}.} It is straighforward to check that the conditions of Theorem 3.1 and Theorem 3.2 of \citet{tjostheim86} are satisfied in our case. Therefore, from these theorems we obtain respectively that $\widehat\alpha$ is strongly consistent and that satisfies the asymptotic normality given in (\ref{an}).

The remaining point that needs to be shown is the expression of the asymptotic variance of $\sqrt{n}(\widehat\alpha-\alpha)$. Following the notation of the paper by \citet{tjostheim86}, we have that $\widetilde Z_{t|t-1}\equiv E(Z_t|Z_{t-1})=\alpha Z_{t-1}+(1-\alpha)\mu$. Hence, we get
\begin{eqnarray*}
&&\hspace{-0.7cm}\frac{\partial \widetilde Z_{t|t-1}}{\partial \alpha}=Z_{t-1}-\mu, \quad U\equiv E\bigg[\bigg(\frac{\partial \widetilde Z_{t|t-1}}{\partial \alpha}\bigg)^2\bigg]=\sigma^2,\\
&&\hspace{-0.7cm} f_{t|t-1}\equiv E[(Z_t-\widetilde Z_{t|t-1})^2|Z_{t-1}]=\mbox{Var}(Z_t|Z_{t-1})=\sigma_\varepsilon^2+\alpha(1+\alpha)|Z_{t-1}|+\frac{2\alpha(1+\alpha)\mu_1\mu_2}{1+\mu_1+\mu_2},\\
&&\hspace{-0.7cm}R\equiv E\bigg[\bigg(\frac{\partial \widetilde Z_{t|t-1}}{\partial \alpha}\bigg)^2f_{t|t-1}\bigg]\\
&&\hspace{-0.7cm}=\sigma^2\bigg(\sigma^2_\varepsilon+\frac{2\alpha(1+\alpha)\mu_1\mu_2}{1+\mu_1+\mu_2}\bigg)+\alpha(\alpha+1)[E(|Z|^3)-2\mu E(sgn(Z)Z^2)+\mu^2E(|Z|)].
\end{eqnarray*}

From the Theorem 3.2 of \citet{tjostheim86}, we obtain that the asymptotic variance of $\sqrt{n}(\widehat\alpha-\alpha)$ is given by $R/U^2$. Using the expressions of $R$ and $U$ above, we obtain the desired result. $\square$\\

We now move our attention for the estimation of $\mu_1$ and $\mu_2$. As mentioned before, the conditional least square method does not provide estimators for these parameters, only for $\mu=\mu_1-\mu_2$. To estimate $\mu_1$ and $\mu_2$ we here propose the method of moments based on the sample quantities of $E(Z)$ and $E(|Z|)$. Before to present explicitly the estimators, we introduce some notation that appears in \citet{koin06} and that will be important for what follows.

Define two real functions $F_1(\cdot,\cdot)$ and $F_2(\cdot,\cdot)$  by
\begin{eqnarray*}\label{F1}
F_1(x,y)=\frac{2y+(x-y)(1+\sqrt{1+4xy})}{(1+\sqrt{1+4xy})(1+x-y)}
\end{eqnarray*}
and
\begin{eqnarray*}\label{F2}
F_2(x,y)=\frac{2y(1+x-y)}{1+2y(x-y)+\sqrt{1+4xy}}.
\end{eqnarray*}

With these definitions above, we immediately obtain from the proof of Proposition 5.2 of \citet{koin06} that the estimators $\widehat\mu_1$ and $\widehat\mu_2$ of $\mu_1$ and $\mu_2$ (respectively) based on the method of moments (with the sample quantities of $E(Z)$ and $E(|Z|)$) are given by
\begin{eqnarray}\label{estmu12_1}
\widehat\mu_1=\frac{F_1(\bar Z_n^+,\bar Z_n^-)}{1-F_1(\bar Z_n^+,\bar Z_n^-)}\quad\mbox{and}\quad \widehat\mu_2=\frac{F_2(\bar Z_n^+,\bar Z_n^-)}{1-F_2(\bar Z_n^+,\bar Z_n^-)},
\end{eqnarray}
if $\bar Z_n\geq0$, and
\begin{eqnarray}\label{estmu12_2}
\widehat\mu_1=\frac{F_2(\bar Z_n^-,\bar Z_n^+)}{1-F_2(\bar Z_n^-,\bar Z_n^+)}\quad\mbox{and}\quad \widehat\mu_2=\frac{F_1(\bar Z_n^-,\bar Z_n^+)}{1-F_1(\bar Z_n^-,\bar Z_n^+)},
\end{eqnarray}
if $\bar Z_n<0$, where $\bar Z_n^+=n^{-1}\sum_{i=1}^n Z_i^+$ and $\bar Z_n^-=n^{-1}\sum_{i=1}^n Z_i^-$. For $x\in\mathbb R$, we have defined $x^+=\max(0,x)$ and $x^-=(-x)^+$.

We now present a proposition that deals with the asymptotic properties of the proposed estimators $\widehat\mu_1$ and $\widehat\mu_2$.

\begin{proposition}
The estimators $\widehat\mu_1$ and $\widehat\mu_2$ given in (\ref{estmu12_1}) and (\ref{estmu12_2}) are strongly consistent for estimating $\mu_1$ and $\mu_2$, respectively, and satisfy the asymptotic normality
$$\sqrt{n}\{(\widehat\mu_1,\widehat\mu_2)-(\mu_1,\mu_2)\}\stackrel{d}{\longrightarrow}{\mbox N}((0,0),\Sigma),$$
as $n\rightarrow\infty$, where the asymptotic covariance matrix $\Sigma$ is given by
\begin{eqnarray}\label{covmatrixSigma}
\Sigma=\frac{\mu_1\mu_2(1+\mu_1)(1+\mu_2)}{(1+\mu_1)(1+\mu_2)+\mu_1\mu_2}
\left(
\begin{array}{cc}
\dfrac{(1+\mu_1)(1+\mu_2)^2-\mu_1\mu_2^2}{\mu_2(1+\mu_2)} & 1 \\
1 & \dfrac{(1+\mu_2)(1+\mu_1)^2-\mu_2\mu_1^2}{\mu_1(1+\mu_1)}
\end{array}
\right).
\end{eqnarray}
\end{proposition}

\noindent{\bf Proof.} Following the ideas of proof of the Theorem 5.2 from \citet{koin06} and using the Law of Large Numbers and Central Limit Theorem for stationary and ergodic processes (instead of classical limit theorems), the proof of our proposition can be obtained and therefore it is omitted. $\square$

From the proposition above, we obtain that the asymptotic distribution of $\widehat\mu=\widehat\mu_1-\widehat\mu_2$ is given by
\begin{eqnarray*}
\sqrt{n}(\widehat\mu-\mu)\stackrel{d}{\longrightarrow}{\mbox N}(0,\Sigma_{11}+\Sigma_{22}-2\Sigma_{12}),
\end{eqnarray*}
as $n\rightarrow\infty$, where $\Sigma_{ij}$ is the element $(i,j)$ of the matrix $\Sigma$ given in (\ref{covmatrixSigma}).
With this, we can construct a confidence interval for $\mu$ and test the null hypothesis $H_0:\mu_1=\mu_2$ against the alternative
hypothesis $H_1:\mu_1\neq\mu_2$. So, we reject the null hypothesis if the value 0 does not belong to the confidence interval of $\mu$.

\section{Simulation issues}\label{simulationissues}

We here present a small numerical experiment to evaluate the finite-sample performance of the estimators
$\widehat{\alpha}$, $\widehat{\mu}_1$ and $\widehat{\mu}_2$ of $\alpha$, $\mu_1$ and $\mu_2$ proposed in the previous section.
We set the sample sizes $n = 50, 100, 200, 400$ and the values of the parameters $\alpha = 0.1, 0.3, 0.5, 0.7$ and $(\mu_1, \mu_2) = (3,3), (6,3)$. To evaluate the point estimation of the parameters we consider the empirical mean and mean squared error. Another interest here is to assess the estimation of the second and third moments of the jump process $J_t$. The Monte Carlo simulation experiments were performed using the \texttt{R} programming language; see \text{http://www.r-project.org}. The number of Monte Carlo replications $R$ considered here was $R=5000$.

Tables \ref{tab1} and \ref{tab2} present the empirical mean and mean squared error of the the estimates of the parameters of our model. From the results presented in these tables, we see that $\widehat{\alpha}$, $\widehat{\mu}_1$ and $\widehat{\mu}_2$ are close to the true values of the parameters for the cases considered, which means that the estimators proposed in the previous section can be used effectively for estimation in the STINAR(1) process. We also observe that the bias of the estimates decreases and the mean square errors go to 0 as the sample size $n$ increases, as expected. Further, the sign of the biases is negative in all cases considered.

\begin{table}[!htbp]
\centering
\caption{Empirical means and mean squared errors (in parentheses) of the estimates of the parameters for $(\mu_1,\mu_2)=(3,3)$
and some values of $\alpha$ and $n$.}\label{tab1}
\renewcommand{\arraystretch}{1.3}
\begin{tabular}{lllccc}
\hline
        $n$&$\alpha$&&$\widehat{\alpha}$&$\widehat{\mu}_1$&$\widehat{\mu}_2$\\ \hline
50         &0.1     &&0.0771 (0.0198)& 2.9556 (0.3924)& 2.9520 (0.3872)                    \\
           &0.3     &&0.2616 (0.0205)& 2.9519 (0.5536)& 2.9369 (0.5457)                     \\
           &0.5     &&0.4444 (0.0223)& 2.9133 (0.8382)& 2.9087 (0.8318)                      \\
           &0.7     &&0.6189 (0.0240)& 2.8572 (1.4046)& 2.7989 (1.5156)                      \\\\
100        &0.1     &&0.0858 (0.0097)& 2.9755 (0.1970)& 2.9709 (0.2015)                     \\
           &0.3     &&0.2792 (0.0104)& 2.9706 (0.2732)& 2.9601 (0.2653)                      \\
           &0.5     &&0.4743 (0.0098)& 2.9575 (0.4253)& 2.9562 (0.4195)                       \\
           &0.7     &&0.6559 (0.0106)& 2.8920 (0.7753)& 2.9078 (0.7652)                      \\\\
200        &0.1     &&0.0932 (0.0053)& 2.9848 (0.0971)& 2.9942 (0.0963)                     \\
           &0.3     &&0.2906 (0.0051)& 2.9864 (0.1342)& 2.9860 (0.1325)                      \\
           &0.5     &&0.4864 (0.0050)& 2.9787 (0.2069)& 2.9695 (0.2108)                       \\
           &0.7     &&0.6773 (0.0047)& 2.9390 (0.3921)& 2.9549 (0.3933)                      \\\\
400        &0.1     &&0.0961 (0.0026)& 2.9971 (0.0500)& 2.9936 (0.0493)                     \\
           &0.3     &&0.2943 (0.0026)& 2.9856 (0.0686)& 2.9901 (0.0694)                      \\
           &0.5     &&0.4927 (0.0024)& 2.9918 (0.1077)& 2.9839 (0.1077)                       \\
           &0.7     &&0.6883 (0.0022)& 2.9796 (0.2064)& 2.9754 (0.2075)                      \\
\hline
\end{tabular}
\end{table}

\begin{table}[!htbp]
\centering
\caption{Empirical means and mean squared errors (in parentheses) of the estimates of the parameters for $(\mu_1,\mu_2)=(6,3)$
and some values of $\alpha$ and $n$.}\label{tab2}
\renewcommand{\arraystretch}{1.3}
\begin{tabular}{lllccc}
\hline
        $n$&$\alpha$&&$\widehat{\alpha}$&$\widehat{\mu}_1$&$\widehat{\mu}_2$\\ \hline
50         &0.1     &&0.0719 (0.0196)& 5.9243 (1.2138)& 2.9433 (0.4930)                     \\
           &0.3     &&0.2625 (0.0198)& 5.8934 (1.6233)& 2.8975 (0.6562)                      \\
           &0.5     &&0.4476 (0.0208)& 5.8366 (2.6158)& 2.8655 (1.0251)                       \\
           &0.7     &&0.6296 (0.0203)& 5.7707 (4.6080)& 2.7778 (1.7914)                      \\\\
100        &0.1     &&0.0848 (0.0099)& 5.9863 (0.6003)& 2.9672 (0.2525)                     \\
           &0.3     &&0.2840 (0.0098)& 5.9630 (0.8015)& 2.9612 (0.3316)                      \\
           &0.5     &&0.4729 (0.0096)& 5.9266 (1.3274)& 2.9115 (0.5132)                       \\
           &0.7     &&0.6586 (0.0094)& 5.8544 (2.4354)& 2.8706 (0.9360)                      \\\\
200        &0.1     &&0.0935 (0.0050)& 5.9867 (0.2980)& 2.9810 (0.1247)                     \\
           &0.3     &&0.2900 (0.0049)& 5.9689 (0.4273)& 2.9839 (0.1707)                      \\
           &0.5     &&0.4843 (0.0046)& 5.9637 (0.6518)& 2.9546 (0.2624)                       \\
           &0.7     &&0.6814 (0.0040)& 5.9321 (1.2533)& 2.9379 (0.4692)                      \\\\
400        &0.1     &&0.0965 (0.0025)& 5.9860 (0.1565)& 2.9900 (0.0646)                     \\
           &0.3     &&0.2940 (0.0025)& 5.9809 (0.2029)& 2.9966 (0.0848)                      \\
           &0.5     &&0.4930 (0.0023)& 5.9793 (0.3321)& 2.9797 (0.1298)                       \\
           &0.7     &&0.6907 (0.0019)& 5.9735 (0.6041)& 2.9750 (0.2401)                      \\
\hline
\end{tabular}
\end{table}

Table \ref{tab3} gives us the empirical mean of the estimates of the second and third moments of $J_t$ and their true values for some values of the parameters and $n=50,100,200,400$; the true values of the moments are replicated for all values of $n$ in order to facilitate the comparison between the estimated and true moments. Here we denote $\mu_J^{(2)}=E(J_t^2)$ and $\mu_J^{(3)}=E(J_t^3)$ and their empirical means by $\widehat\mu_J^{(2)}$ and $\widehat\mu_J^{(3)}$, respectively. We see a good performance of the estimated second and third moments of $J_t$ based on the estimators given in the previous section, since they are close to the true values of the moments for all cases considered here. As expected, we observe that the biases decrease
as the sample size increases.

\begin{table}[!htbp]
\centering
\caption{True values and empirical means of the second and third moments of $J_t$ for some values of the parameters and $n=50,100,200,400$.}\label{tab3}
\renewcommand{\arraystretch}{1.3}
\begin{tabular}{ll|llllr|llllll}
\hline$(\mu_1, \mu_2)$  &        &$(3,3)$  &&&             &&&$(6,3)$&&&&\\ 
\hline
$n$&$\alpha$&$\mu_J^{(2)}$&$\widehat{\mu}_J^{(2)}$&&$\mu_J^{(3)}$&$\widehat{\mu}_J^{(3)}$&&$\mu_J^{(2)}$&$\widehat{\mu}_J^{(2)}$&&$\mu_J^{(3)}$&$\widehat{\mu}_J^{(3)}$  \\  
\hline
50         &0.1     &43.2&44.80 &&0&0.396     &&97.2&100.3 && 114.8&72.31 \\
           &0.3     &33.6&35.47 &&0&0.638     && 75.6&79.38 &&255.9&237.5       \\
           &0.5     &24.0&25.92 &&0&$-$0.925  &&54.0&58.02 &&279.0&290.5      \\
           &0.7     &14.4&16.55 &&0&$-$0.418  &&32.4&37.86 &&183.9&254.8       \\\\
100        &0.1     &43.2&44.04 &&0&   0.075  && 97.2&98.88 && 114.8&92.42 \\
           &0.3     &33.6&34.39 &&0&$-$0.552  && 75.6&77.71 &&255.9&247.3   \\
           &0.5     &24.0&24.90 &&0&0.176     &&54.0&56.21 &&279.0&289.8    \\
           &0.7     &14.4&15.49 &&0&0.593     &&32.4&35.46 &&183.9&223.9           \\\\
200        &0.1     &43.2&43.54  &&0&0.069    && 97.2&98.13 && 114.8&105.1   \\
           &0.3     &33.6&33.91  &&0&0.026    && 75.6&76.07 &&255.9&249.9    \\
           &0.5     &24.0&24.47  &&0&$-$0.174 &&54.0&55.48 &&279.0&285.9     \\
           &0.7     &14.4&15.00  &&0&0.217    &&32.4&33.75 &&183.9&201.8     \\\\
400        &0.1     &43.2&43.41  &&0&0.048 && 97.2&97.79 && 114.8&108.2   \\
           &0.3     &33.6&33.71  &&0&0.040 && 75.6&75.87 &&255.9&254.4     \\
           &0.5     &24.0&24.23  &&0&0.167 &&54.0&54.71 &&279.0&280.7    \\
           &0.7     &14.4&14.73  &&0&0.193 &&32.4&33.20 &&183.9&193.2    \\
\hline
\end{tabular}
\end{table}

\section{Application}\label{application}

We here show the usefulness of the STINAR(1) process by applying it to a real data set.
We consider the time series of annual Swedish population increases (per
thousand population) for the 1750--1849 century as reported in \citet{thomas40}
denoted by $Z_t$, which is presented in Table \ref{data} and can be also obtained online at the site http://robjhyndman.com/TSDL. 
This data set was used recently in \citet{ky09} and \citet{kt11}.

\begin{table}[!htb]
\begin{center}\caption{The Swedish population rates series (in 1000s) from 1750 to 1849.}\label{data}
\begin{tabular}{rrrrrrrrrrrrrrrrr}
\hline
 9  &12   &8  &12  &10  &10   &8      &2   &0   &7    &10   &9   &4   &1    &7   &5   &8   \\
 9   &5   &5  &6   & 4  &$-$9 &$-$27  &12  &10  &10   &8    &8   &9   &14   &7   &4   &1   \\
 1   &2   &6   &7   &7  &$-$2 &$-$1   &7   &12  &10   &10   &4   &9   &10   &9   &5   &4   \\
 3   &7   &7   &6   &8  &3    &4      &$-$5 &$-$14& 1 &6    &3   &2   &6    &1   &13  &10  \\
 10  &6   &9  &10   &13 &16   &14     &16   &12   &8  &7    &6   &9   &4    &7   &12  &8   \\
 14  &11  &5   &5   &5  &10   &11     &11   &9   &12  &13   &8   &6  &10  &13\\
\hline
\end{tabular}
\end{center}
\end{table}

\begin{figure}[!h]
\centering \caption{Plots of the time series, autocorrelation and partial autocorrelation functions for the Swedish population rates series (in 1000s) from 1750 to 1849.}
		 \includegraphics[width=0.9\textwidth]{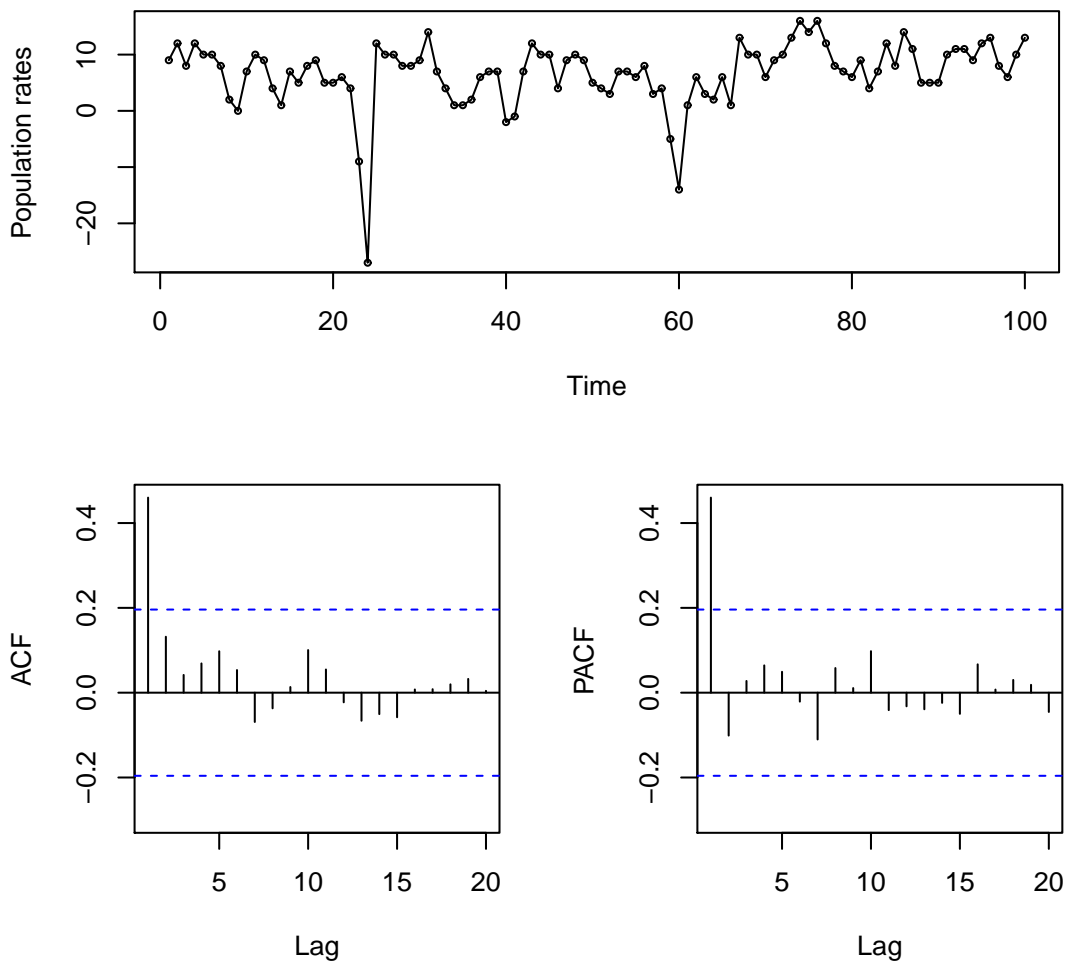}
\label{fig9}
\end{figure}

Table \ref{ed} displays some descriptive statistics of the Swedish population rates series. We see that the series contains negative integer values and therefore the usual count time series models can not be applied in this case. The time series data and their sample autocorrelation and partial autocorrelation are displayed in the Figure \ref{fig9}.
\begin{table}[!htb]
\begin{center}\caption{Descriptive statistics for the Swedish population rates series (in 1000s) from 1750 to 1849.}\label{ed}
\label{tableaplic1}
\begin{tabular}{cccccc}
\hline
Minimum  & Median & Mean & Variance &$\widehat{\rho}(1)$ &Maximum \\ \hline
$-$27.00 &7.50    &6.69  & 34.56    &0.46                &16.00   \\ \hline
\end{tabular}
\end{center}
\end{table}

Figure \ref{fig9} suggests that a first-order autoregressive model may be appropriate for fitting the time series considered here since the sample autocorrelations presents a geometric decay (as the lag increases) and the partial autocorrelations have a clear cut-off after lag 1; \citet{ky09} and \citet{kt11} also proposed AR(1) processes to fit this data set. Furthermore, the behavior of the series indicates that it can be generated by a mean stationary model. 

We here also compare our STINAR(1) with the TINAR(1) introduced by \citet{free10}. In order to make a fair comparison, we here consider an asymmetric version of the TINAR(1) process as discussed in Section 6 of \cite{free10}. With this, the TINAR(1) model considered here has marginals following a Skellam distribution with parameters $\lambda_1(1-\beta)^{-1}$ and $\lambda_2(1-\beta)^{-1}$ ($\lambda_1,\lambda_2>0$ and $0<\beta<1$), that is, the marginals are distributed as $Y_1-Y_2$, where $Y_1$ and $Y_2$ are two independent Poisson random variables with mean $\lambda_1(1-\beta)^{-1}$ and $\lambda_2(1-\beta)^{-1}$, respectively, and $\beta$ is the associated thinning parameter of this process. To estimate $\lambda_1$ and $\lambda_2$, we use the sample quantities of $E(Z_t)$ and $\textrm{Var}(Z_t)$, which in the asymmetric version of the Freeland model are given by $E(Z_t)=(\lambda_1- \lambda_2)(1-\beta)^{-1}$ and $\textrm{Var}(Z_t)=(\lambda_1+\lambda_2)(1-\beta)^{-1}$, respectively. We estimate the parameter $\beta$ through the conditional least square method, which yields the same estimator of that proposed here for our thinning parameter $\alpha$.

In the Table \ref{tableaplic1} we present the estimates of the parameters and four goodness of-fit statistics: RM (root mean of differences between observed and predicted values), RMS (root mean square of differences between observed and predicted values), MA (absolute mean of differences between observed and predicted values) and MDA (absolute median of differences between observed and predicted values); here the predicted values are obtained by the estimated conditional expectation $E(Z_t|Z_{t-1})$. In general it is expected that the better model to fit the data presents the smaller values for these quantities. For a good discussion of these statistics, we recommend the reader to the paper by \citet{hyko06}.

\begin{table}[!htbp]
\centering
\caption{Estimate of the parameters and the goodness-of-fit statistics RM, RMS MA and MDA for the STINAR(1) and TINAR(1) processes.}\label{tableaplic1}
\begin{tabular}{lcccccccccc}
 \hline
Model            &&Estimates                         &&RM   &&RMS   &&MA  &&MDA  \\
 \hline
STINAR(1)        &&$\widehat{\alpha} = 0.465$       &&0.0796 &&5.2064 &&3.4200 &&2.4381 \\
                 &&$\widehat{\mu}_1  = 8.883$       &&       &&       &&   &&      \\
                 &&$\widehat{\mu}_2  = 2.193$       &&       &&       &&   &&      \\\\
TINAR(1)         &&$\widehat{\beta} = 0.465$       &&0.0804 &&5.2064 &&3.4201 &&2.4379 \\
                 &&$\widehat{\lambda}_1  = 11.03$   &&       &&       &&   &&     \\
                 &&$\widehat{\lambda}_2  = 7.449$   &&       &&       &&   &&     \\
\hline
\end{tabular}
\end{table}

From the Table \ref{tableaplic1} we see that our STINAR(1) process yields a slightly better fit to the data than the asymmetric version of the \cite{free10} model based on the goodness-of-fit statistics. 

The standard errors for the estimates of the parameters $\alpha$, $\mu_1$ and $\mu_2$ are respectively $0.0955$, $0.9992$ and $0.4364$. The estimated covariance between $\widehat\mu_1$ and $\widehat\mu_2$ is $0.12045$. We also obtain confidence intervals with a significance level at $5\%$ for the parameters $\alpha$, $\mu_1$ and $\mu_2$, which are given by $(0.2778;0.6522)$, $(6.9246;10.841)$ and $(1.3376;3.0484)$, respectively. In order to test the null hypothesis $H_0: \mu_1=\mu_2$ against the alternative hypothesis $H_1: \mu_1\neq\mu_2$, we construct a confidence interval (at a significance level of $5\%$) for $\mu$ as proposed in the final of Section \ref{estimationandinference}. The confidence interval for $\mu$ is $(4.7817; 8.5983)$ and since it does not contain the value 0 we reject the null hypothesis in favor of the alternative hypothesis that states that the data were generated by a STINAR(1) model with $\mu_1\neq\mu_2$.

\begin{figure}[!h]
\centering \caption{Plots of the sample autocorrelations of the residuals and the jumps against time.}
         \includegraphics[width=0.45\textwidth]{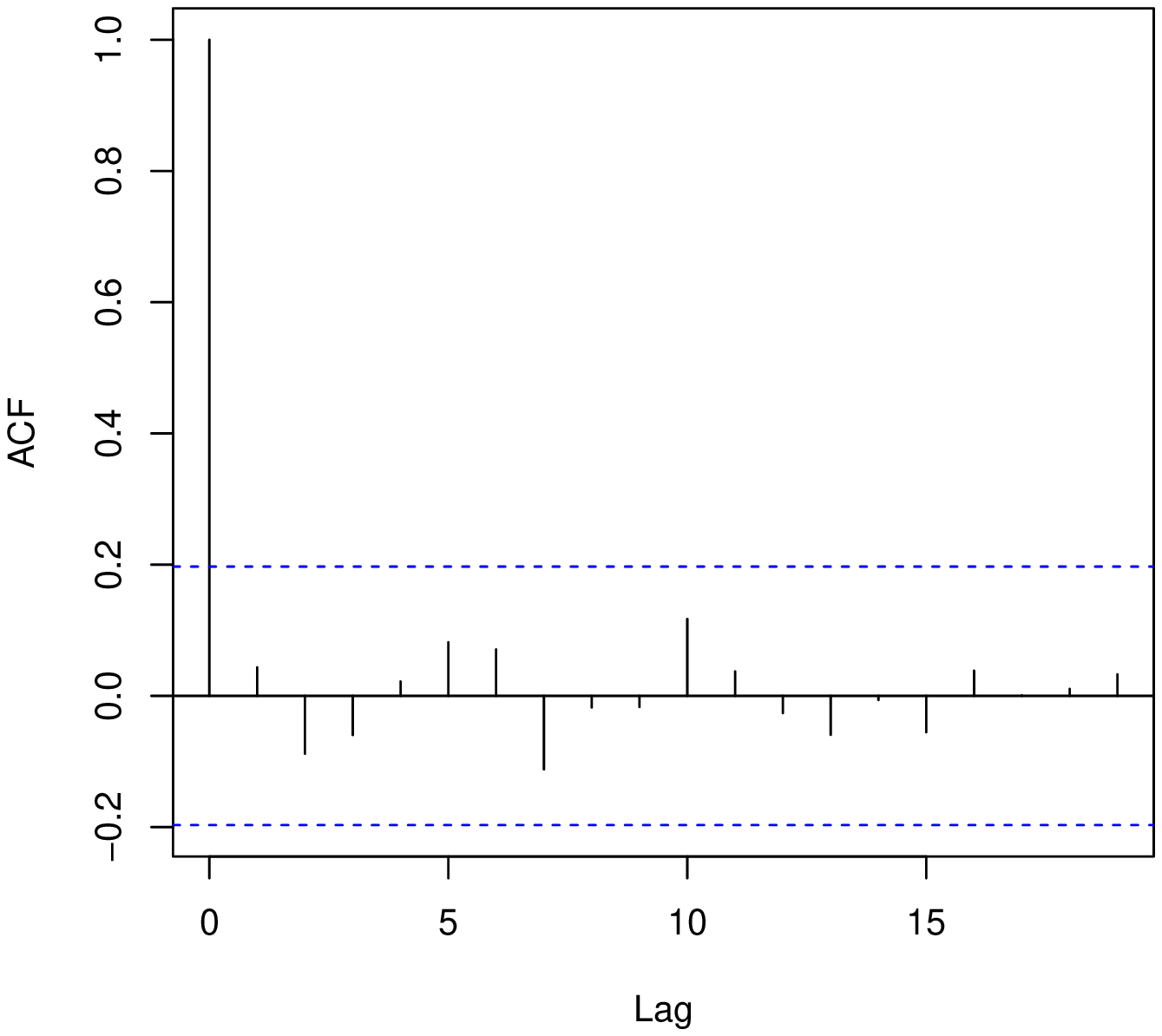}
         \includegraphics[width=0.45\textwidth]{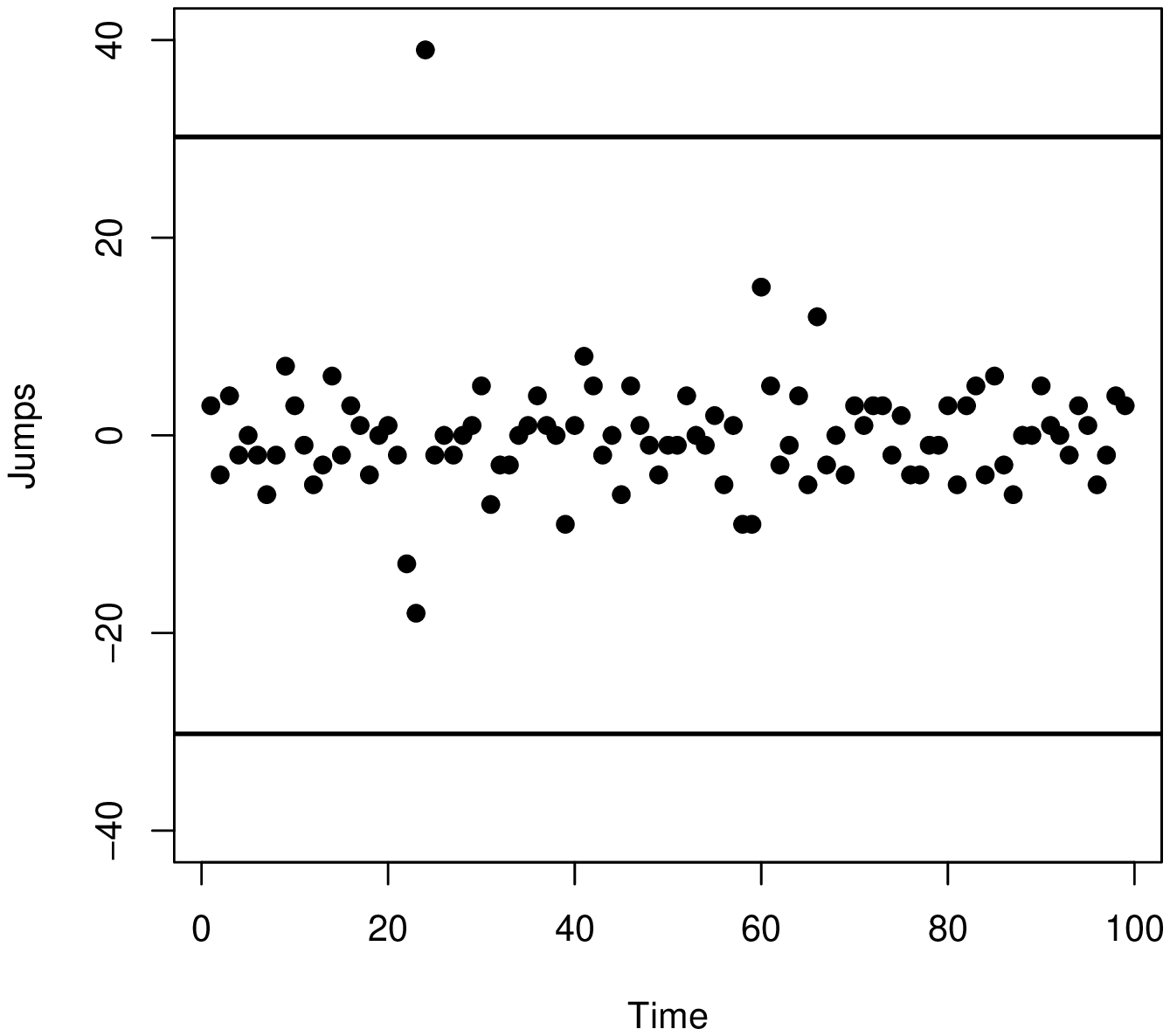}
\label{fig10}
\end{figure}

In the Figure \ref{fig10} we present plots of the sample autocorrelations of the ordinary residuals and the jumps against time with $\pm3\sigma_J$ limits chosen as the benchmark chart as proposed by \cite{weib09b}; here we define $\sigma_J\equiv\sqrt{\mbox{Var}(J_t)}$, where the variance of $J_t$ is given in Proposition \ref{propmomjumps}. These plots indicate that the residuals are not correlated and that our AR(1) model is well fitted.

\section*{Acknowledgements}

The authors thank the financial support from {\it Conselho Nacional de Desenvolvimento Cient\'ifico e Tecnol\'ogico} (CNPq-Brazil) and {\it Coordena\c c\~ao de Aperfei\c coamento de Pessoal de N\'ivel Superior} (CAPES-Brazil).

\end{document}